# Towards scalable silicon quantum computing

M. Vinet[1*], L. Hutin[1], B. Bertrand[1], S. Barraud[1], J.-M. Hartmann[1], Y.-J. Kim[1], V. Mazzocchi[1], A. Amisse[1,2], H. Bohuslavskyi[1,2], L. Bourdet[2], A. Crippa[2], X. Jehl[2], R. Maurand[2], Y.-M. Niquet[2], M. Sanquer[2], B. Venitucci[2], B. Jadot[3], E. Chanrion[3], P.-A. Mortemousque[3], C. Spence[3], M. Urdampilleta[3], S. De Franceschi[2] and T. Meunier[3]

Université Grenoble Alpes, France
[1] CEA, LETI, Minatec Campus, F-38054 Grenoble, [2] CEA, INAC, F-38054 Grenoble, [3] CNRS, Institut Néel, F-38042 Grenoble
*e-mail: maud.vinet@cea.fr / Phone: (+33) 438 78 99 00

*Abstract*— We report the efforts and challenges dedicated towards building a scalable quantum computer based on Si spin qubits. We review the advantages of relying on devices fabricated in a thin film technology as their properties can be in situ tuned by the back gate voltage, which prefigures tuning capabilities in scalable qubits architectures.

## I. INTRODUCTION

It is now well acknowledged that quantum computing (QC) will extend high performance computing roadmap [1-2]. However, to be a serious contender to classical computers, digital QC will have to perform large number of quantum operations and thus manipulate large numbers of quantum bits. Typically, to solve problems beyond classical computer reach, quantum operations will be over a billion. To cure errors due to decoherence, quantum error correction techniques, which utilize the idea of redundant encoding, are needed to define logical qubits on which those operations are performed [3-5]. With state-of-the-art codes, error thresholds or fidelities (around $10^{-2}$ in Si spin qubits), it is believed that logical qubits will be made out of a few thousands or more of physical qubits [6] bringing the number of required physical qubits to perform relevant quantum calculations to at least a million.

Because of these large numbers of operations and physical qubits, Si-based QC appears as a promising approach due to the size of the qubits, the quality of the quantum gates and the VLSI ability to fabricate billions of closely identical objects. However, silicon spin qubits have a quite recent history. The first single qubit relying on a confined electron spin in silicon were only reported in 2012 [7]. Since then, their development in terms of gate realization has been comparatively fast. Moreover, thanks to the introduction of isotopically purified $^{28}$Si, a large enhancement of spin coherence has been observed. Multiple research groups have realized single- and two-qubit quantum gates with already high and yet improving fidelities, see **fig.1** [8-11]. Recently, first qubit functionalities in a device fabricated in a 300mm CMOS platform have been demonstrated [12, 13]. All these results provide encouraging building blocks for scalable, fault-tolerant QC.

In this paper, we will discuss the figures of merit and review the challenges to be tackled to build a large-scale QC. We will show how FDSOI technology features can be leveraged for quantum chip optimization.

## II. EFFORT TOWARDS RELIABLE AND REPRODUCIBLE SINGLE AND TWO QUBITS GATES

There is no consensus yet on the most relevant figures of merit and the required trade-offs optimizing quantum dot and qubit variability and fidelity mostly due to the lack of statistical data. Efforts are simultaneously made in several directions.

### A. Material optimization

First from a pure material perspective, due to the absence of residual nuclear spins, isotopically purified $^{28}$Si has already shown very promising improvement in spin coherence times [11]. We have grown $^{28}$silicon epitaxial layers with an isotopic purity greater than 99.992 % on 300mm natural abundance silicon crystalline wafers. The quality of the mono-crystalline $^{28}$Si epilayer respects the same drastic quality requirements as the natural epilayers used in standard CMOS technology. Synthesis and enrichment of SiF$_4$ took place in Russia at SC "PA Electrochemical Plant" (ECP). The synthesis of silane was carried out by the reaction of high-purity silicon tetrafluoride with calcium hydride. We profiled with secondary ion mass spectrometry (SIMS) the various Si isotopes in a 60 nm thick Si layer grown at 650°C, 20 Torr with $^{28}$SiH$_4$ on a natural abundance Si(001) wafer (**fig. 2**) [15].

Then to reduce potential fluctuations in CMOS and ensure that quantum dots are actually gate-defined rather than by disorder, surface roughness and gate stack will have to be carefully engineered. Several calculations attempted to quantify the impact of physical dimensions on potential fluctuations have been performed [15, 16]. For qubits defined by nanowires in thin film devices, it means that both thickness and etched defined edges need to be as smooth as possible as presented in **Fig. 3**. **Fig. 4** quantifies surface roughness depending on several finishing preparations showing that Rms can be brought down to 0.1nm range [17]. Qubits will harness developments towards reducing $D_{it}$ or gate stack granularity as well.

### B. Hole and electrons CMOS compatible qubits

The first qubit implemented on a foundry-compatible Si CMOS platform was built using a SOI NanoWire MOSFET, it is in essence a compact two-gate pFET, **fig. 5** [12]. The inhomogeneous dephasing time $T_2^*$ was limited to 60ns as the studied EDSR transition was involving an excited orbital state. In a recent experiment on a similar device we investigated the coherence of an EDSR transition between spin-orbit states within same orbital. It revealed an extended $T_2^*$=270ns close to the inhomogeneous dephasing time of electron in natural Si.

In addition, in SOI architecture, the back bias ($V_{bb}$) can be leveraged to in-situ tune the system properties. First we have evidenced that the inter-dot coupling, mediated by tunneling and Coulomb interaction, can be tuned over 6 orders of magnitude by means of $V_{bb}$. [18], **fig 6.** Then, the Rabi frequencies of qubits show complex dependence on $V_{bb}$, which results from the control of the shape and symmetry of the wave functions. The qubits may, therefore, be switched between bias points where they can be efficiently manipulated electrically and others where they are far less responsive but (as a consequence) decoupled from the electric noise and longer-lived.

In the case of electrons, the spin-valley mixing has led to the experimental observation of electrically-induced spin resonance in Si without having to co-integrate micro-magnets [19]. As for holes, using $V_{bb}$ to tune the vertical confinement (and thus valley splitting) enables continuously switching between a protected spin regime and an E-field addressable valley regime [20, 21]. Trade-offs for optimization still need to be found to take advantage of this in situ tuning capability in terms of number of operations per error or architecture.

### C. Readout measurements

We have recently shown [22] an ultra-compact device fabricated in foundry-compatible Si MOS technology, with a built-in charge detector (SET) capacitively coupled to two Gate-defined QDs. Thanks to an energy-selective detection scheme, we have demonstrated single-shot readout of a two-electron spin-state in a "corner QD" by measuring the time trace of the SET current. As detailed in **Fig. 7**, the optimization of the readout speed/fidelity trade-off in this readout scheme is carried out by increasing the tunnel rates while keeping a large ($\Gamma_{SET}$-$\Gamma_{QD}$) window. Here again, $V_{bb}$ was used to tune the cross-capacitance between the SET and the QD in order to improve the readout signal and shorten the minimum integration time.

In this geometry, we have further investigated the quality of the readout that can be estimated by measuring the amplitude of the detector signal in the two different spin configurations, as presented by the histogram in **fig. 8**. We extract the so-called readout fidelity which is as good as 99.9% for 1ms of integration time. Alternative methods such as gate-reflectometry are under investigation, it is considered a more compact and scalable readout method. In this technique, the charge sensing required to sense the qubit state is accomplished by measuring the dispersive response of an electromagnetic RF resonator connected to one of the qubit gates and excited at its resonance frequency. We have obtained the complete charge stability diagram of a coupled two-dot system [23].

### D. Possible manipulation schemes

A first way of driving coherent rotation of a spin is through Electron Spin Resonance, or ESR. Experimentally, one can deposit in close proximity of the device a microstrip line that is used to flow a large AC current and generate an oscillating magnetic field resonant with the spin transition frequency. Coupling the spin to an RF magnetic field seems like the most straightforward method, although the excitation is hardly applied locally. This can be a drawback for maximizing the manipulation speed which depends on the coupling strength and is typically in the range of 1MHz for this scheme [8].

A second mechanism is the Electric Dipole Spin Resonance (EDSR). In this case, the spin rotation is induced by an oscillating electric field, which can be provided by a field-effect Gate placed directly above the QD. If the properties of the system are such that Spin-Orbit Coupling (SOC) is significant, the orbital motion caused by an RF E-Field alone can drive spin rotations [12, 13]. Otherwise a possible approach consists in embedding a micro-magnet as an auxiliary in the vicinity of the device, causing the particle traveling back and forth to perceive an oscillating B-field [9-11]. Although efficient for fast manipulation of a few qubits, this technique may become problematic for the design and integration at large-scale. Note that for both implementations of EDSR, a stronger coupling can lead to ~100MHz spin rotations, but sensitivity to charge noise may limit the coherence time (**fig. 1**).

## III. QUBITS IN AN ARRAY OF QUANTUM DOTS

### A. Qudots definition and coupling to the nearest neighbours

The challenge towards large-scale integration comes down to the ability to individually control single spin qubits in arrays of millions of quantum dots, together with controlling the nearest-neighbor interaction. By resorting to a line/column addressing in a 2D architecture and a definition of the dots through the potential applied to their surrounding tunnel barriers (**fig. 9**), the number of gates can be scaled down proportionally to sqrt(N). In this layout scheme, individual quantum dots as well as the interaction between adjacent dots are defined by 4 gate voltages (**fig. 10**).

### B. Cell for full quantum functionalities

In a CMOS-compatible technology, the only demonstrated qubit so far looks like two gates in series on a nanowire with one dot used as a qubit and the second as its readout or sensing dot [12, 22]. However, this charge detector presents some limitation in terms of scalability as the presence of two reservoirs in series to perform current measurements is required. To extend this principle and design a 2D array required for quantum error correction, we propose to locate the sensing dot in a layer below the qubit in order to build a compact and local unit-cell device containing a spin qubit with all the quantum functionalities (**fig. 11**). The top layer is used to encode the quantum information and the bottom layer is used to design local electrometers for readout, **fig. 12**. By using a spin-to-charge conversion and applying RF reflectometry techniques to the source of the detector, it provides a compact and scalable qubit & read-out design. The bottom layer is also connected to electron reservoirs to allow a scalable initialization to overcome the challenges of loading electrons from the sides of the array [24].

## IV. QUANTUM CHIPS ARCHITECTURE

Because of Zeeman splitting, Si spin computing is limited to very low temperature operation. It impose very stringent conditions on power consumption for classical control electronics surrounding the quantum core, **fig. 14**. In view of their co-integration, we have demonstrated that unique back-biasing capability to FDSOI technology can be used at low temperature to compensate for Vt increase as shown in **fig. 15** and dynamically tune the performance/consumption trade-off [25].

Finally, we have demonstrated that package optimization through thermal coating or insertion of heat management modules can dramatically improve heat dissipation [26].

## V. CONCLUSION

Electron and hole qubits have been fabricated in SOI nanowire-like integration. We have shown that the back gate voltage can be used to in situ adjust the system properties; i) to tune the tunneling coupling between dots over 6 orders of magnitude; ii) to optimize the readout speed/fidelity trade-off, we used it to demonstrate a read-out fidelity of 99.9% for 1ms of integration time; iii) to enable continuous switching between a protected regime and an E-field addressable regime. To upscale the number of qubits, we have proposed a vertical integration with sensing dots located below the qubits to enable a compact 2D array of qubits with full quantum functionalities.

| ref | techno | $T_\pi$ | $T_2^*$ | $Q = \dfrac{T_2^*}{T_\pi}$ | 1-qubit gate fidelity | 2-qubit gate fidelity |
|---|---|---|---|---|---|---|
| Yoneda2018 (RIKEN) | $^{28}$Si/SiGe µ-magnet | 25 – 100 ns | 20 µs | 200 – 800 | > 99.9 % | |
| Watson2018 (QuTech) | Si/SiGe µ-magnet | 250 ns | 0.6 – 1 µs | 4 | 98 – 99 % | ~ 90 % ~ 800 ns CNOT |
| Zajac2018 (Princeton) | Si/SiGe µ-magnet | 100 ns | 1.2 – 1.4 µs | 12 | > 99 % | ~ 90 % ~ 200 ns CNOT |
| Chan2018 (UNSW) | $^{28}$Si bulk ESR antenna | 1 µs | 30 µs | 30 | ~ 99.9 % | |
| Huang2018 (UNSW) | $^{28}$Si bulk ESR antenna | 1 µs | 10 – 30µs | 10 – 30 | | ~ 90 % ~ 1 µs CROT |
| Maurand (CEA) | FDSOI – hole qubit - EDSR | 5 ns | 0.3 µs | 60 | | |

Fig. 1: Table summarized the Si spin qubits state of the art.

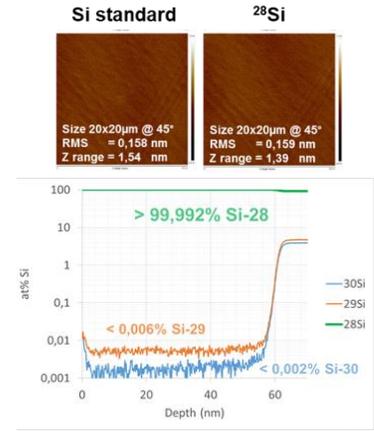

Fig. 2: Top - 20x20µm² AFM scan of standard Si surface compared to epitaxially grown $^{28}$Si. Both surfaces exhibits same surface roughness. Bottom: SIMS characterization of $^{28}$Si after growth showing about 99.992% of $^{28}$Si.

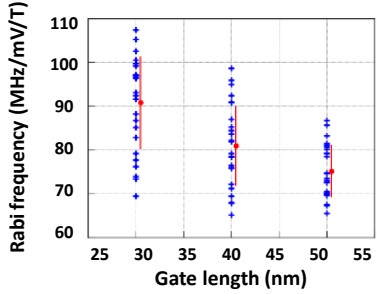

Fig. 3: Rabi frequency in rough hole qubit as a function of gate length. Each cross is a different realization of a Gaussian surface roughness profile with rms = 0.4 nm. The red dot and bar are mean and standard deviation.

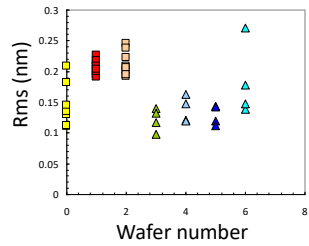

Fig. 4: SOI surface roughness illustrating that it can be decreased thanks to wafer fabrication process tuning.

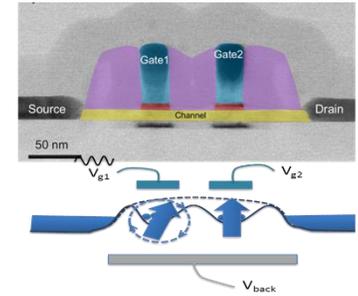

Fig. 5: First qubit implemented on a foundry-compatible Si CMOS platform. It consists of a SOI NanoWire MOSFET with two-gate pFET in a 64nm pitch.

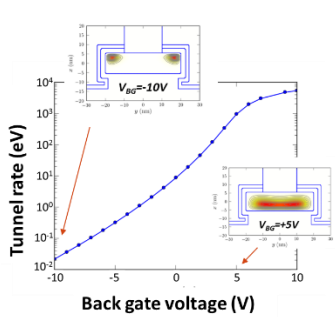

Fig. 6: A two-band k.p model accounting for valley-orbit coupling was used to calculate single-electron states. The tunnel coupling t was extracted versus $V_{BG}$ from the anticrossing between the lowest single-electron states. This shows the tunability of inter-dot transport by the Back-Gate voltage.

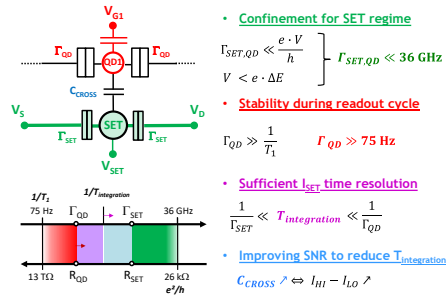

Fig. 7: Readout by transport through a coupled SET charge detector. The charge events in QD1 are assumed to be spin-dependent. Design windows in terms of Γ tunnel rates, and corresponding tunnel resistance are represented (V=150µV, and $T_1$=13.5ms). Increasing the red area yields faster readout. Maximizing the SET-Dot $C_{cross}$ leads to improving the readout signal and thus enables to reduce the integration time, i.e. squeeze the blue area. $V_{bb}$ can be a handle for tuning $C_{CROSS}$.

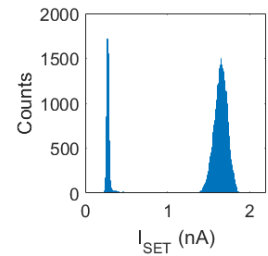

Fig. 8: Histogram of the level of current of the single-shot traces for the singlet and triplet spin states after 1ms integration time. The detection bandwidth is fixed by room-T° electronics to 1 kHz. A spin read-out fidelity above 99.9% is achieved and the relaxation was measured close to 13.5 ms.

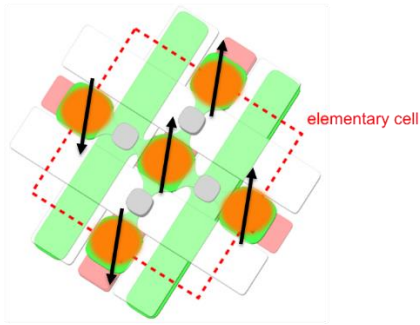
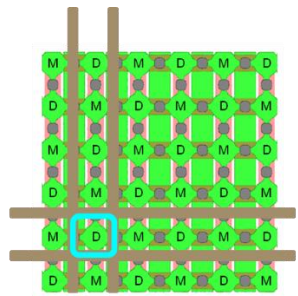
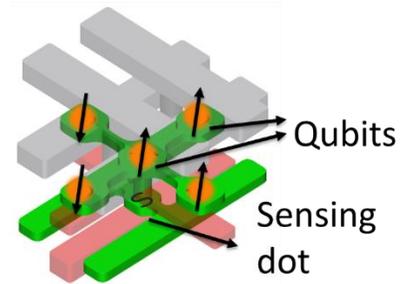

Fig. 9: Unit 2D cell where the cental dot is defined thanks to voltage applied through the grey gates on the constrictions with its nearest neighbors.

Fig. 10: 2D array with one dot selected thanks to gate bias applied to its 4 neighbouring tunnel junctions.

Fig11: Vertical unit cell with the sensing dot located below its qubits and coupled through a controllable and addressable tunnel junction.

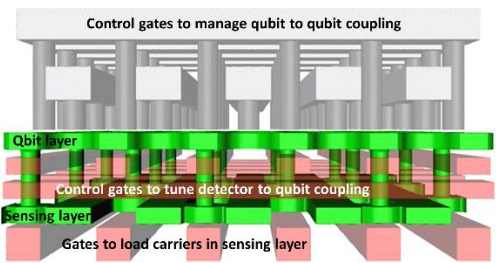
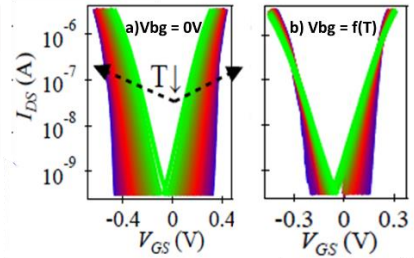

| T° of operation | Typical cooling power |
|---|---|
| 20mK | 30μW |
| 100mK | 1mW |
| 1K | 1W |

Fig. 12: The qubit contains the quantum hardware with spin qubits stored in an array of tunnel-coupled quantum dots. To control the quantum dots, a grid of long gates is designed and allows individual electrostatic and coherent control of the electron spin qubits. The lower layer is dedicated to the engineering of local reservoirs of electrons and local electrometers for the qubit layer. The two layers are coupled by a controllable tunnel barrier to allow electron transfer between them. This tunable tunnel coupling will be used to perform the read-out and the initialization of the quantum hardware.

Fig. 13: Cooling power table as a function of T(K).

Fig. 14: Close-up in the threshold voltage region of Id versus gate voltage curves for temperature ranging from 300 (green to blue shades) down to 4K. Left panel shows Vt increase around 0.5V both for N and PMOS that can be corrected thanks to Vbg as shown in right panel.

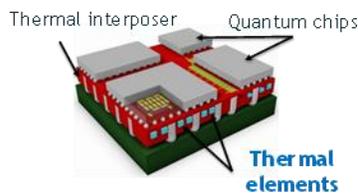

Fig. 15: Thermal elements can be used both as thermal insulator to force a gradient between the coldest part of the core quantum chip and some parts of classical electronics control. They can also be used to manage heat dissipation by damping heat burst into phase change materials.

**ACKNOWLEDGMENT -** The authors gratefully acknowledge financial support from the EU under Project MOS-QUITO (No. 688539) and the Marie Curie Fellowship within the Horizon 2020 program and from French Agence Nationale de la Recherche through the projects ANR-15-IDEX-02 and the ANR-16-ACHN-0029.